\documentclass[journal]{IEEEtran}

\usepackage{cite}

\ifCLASSINFOpdf
  \usepackage[pdftex]{graphicx}
  \graphicspath{{.}}
  \DeclareGraphicsExtensions{.eps,.jpg}
\else
  \usepackage[dvips]{graphicx}
  \graphicspath{{.}}
  \DeclareGraphicsExtensions{.eps,.jpg}
\fi

\usepackage{amsmath,amssymb}
\usepackage{array}
\usepackage[hidelinks]{hyperref}

\begin{document}

\title{Terahertz frequency selective surfaces using\\ heterostructures based on two-dimensional\\ diffraction grating of single-walled\\ carbon nanotubes}

\author{Alexander~M.~Lerer,
	Pavel~E.~Timoshenko
	and~Sergei B. Rochal% <-this % stops a space
	\thanks{A.\,M.~Lerer, P.\,E.~Timoshenko, and S\,B.~Rochal are with the Faculty of Physics, Southern Federal University, Rostov-on-Don, Russia 344090, email:~\href{mailto:p.e.timoshenko@gmail.com}{P.E.Timoshenko@GMail.com}}}

\markboth{arxiv.org, 9~March~2022}%
{A.\,M.~Lerer \MakeLowercase{\textit{et al.}}: FSS using heterostructures based on 2D grating of SWCNTs}

\maketitle

\begin{abstract}
For single-walled carbon nanotubes (SWCNTs) with a length of 1--50~$\mathbf{\mu m}$, the surface plasmon-polariton (SPP) resonance is within the terahertz frequency range; therefore, SWCNT lattices can be used to design frequency-selective surface (FSS). The numerical model of electromagnetic wave diffraction on a two-dimensional periodic SWCNT lattice can be described by an integro-differential equation of the second-order with respect to the surface current along SWCNT. The equation can be solved by the Bubnov-Galerkin method. Frequency dependence of reflecting and transmitting electromagnetic waves for FSSs near the SPP resonance are studied numerically. It is shown that the resonances are within the lower-frequency part of the terahertz range. Also,  we estimate the relaxation frequency of an individual SWCNT and demonstrate the applicability of the Kubo formula for graphene conductivity to array of strips similar in size with SWCNTs under consideration.
\end{abstract}


% \fi
\IEEEpeerreviewmaketitle

\section{Introduction}\label{sec:intro}
\IEEEPARstart{I}n 2021, it has been 30 years since the beginning of the synthesis and study of the properties of carbon nanotubes. During the period opened by Sumio Iijima~\cite{Iijima1991}, these nanostructures have become extremely attractive for state-of-the-art technology due to their mechanical, chemical and electrodynamic properties.

Single-walled carbon nanotube (SWCNT) is a seamless cylinder that can be represented as a result of rolling up of a flat hexagonal graphene strip.~\cite{Saito1998, Jorio2008} The SWCNT diameter ranges from one to several tens of nanometers, whereas the length can reach a millimeter size. Orientation of the honeycomb graphene lattice with respect to nanotube axis determines an important structural characteristic, i.e. chirality. More precisely, the chirality is characterized by two integer indices $(m, n)$ known as chiral vector and indicating the location of the lattice hexagon which, as a result of the tube rolling up, shall coincide with the hexagon at the origin.~\cite{Dresselhaus1995} If the difference of the chirality indices is divisible by 3, the nanotube is metallic, otherwise, it is semiconductor.~\cite{Saito1992, Saito2000} In addition to the type of conductivity, the bandgap that is an important characteristic of electronic properties of semiconducting nanotubes, is also determined by the SWCNT chirality.~\cite{Samsonidze2003}

Carbon nanotubes can be applied to the surface of any shape.~\cite{Li2019,Rao2003,Che1998,KholghiEshkalak2017} This leads to the opportunity to manufacture high-performance frequency-selective surfaces (FSS)~\cite{Matsumoto2015,Cao2009} with desired reflecting and absorbing bands, representing a biperiodic grating of nanotubes located on a layered substrate. Due to the large length-to-diameter ratio and sufficiently high conductivity, electric fields with a high time-averaged strength arise in the resonance region near the relaxation frequency of the nanotube. According to~Ref.~\cite{Melnikov2021}, the resonant frequencies of SWCNTs lie within the lower-frequency part of the sub-terahertz range. Devices operating in this range are widely used in various fields from astrophysics, ecology and environmental monitoring in order to examine high-purity substances, communication and security systems, etc. Due to the promising properties of nanotubes, the size of nanotube-based diffraction lattices is expected to be an order of magnitude smaller than similar gratings based on metal strips.

The design of these devices has a high priority and requires the strictly electrodynamic modeling. However, the study of heterostructures by direct grid methods in the variational (finite element method, method of moments) and differential (finite difference method) formulations encounters a number of difficulties associated with the application of these methods to semi-infinite layers, periodic and open boundaries. Construction of the tensor Green's function for Maxwell's equations in a layered medium and solution of the vector integral equations by the Bubnov-Galerkin projection method is the most rigorous approach for numerical modeling this problem.~\cite{Lerer2012, Lerer2021} In contrast to the latter papers, we note that in the diffraction problems the transverse current in the SWCNT can be neglected.

In this paper, the diffraction of electromagnetic waves on a two-dimensional periodic grating of SWCNTs placed on a multilayered substrate are studied. The substrate can be made up of dielectric, metal, and graphene layers.

In the considered theory of SWCNT diffraction, it was assumed that the following boundary condition is satisfied on the nanotube surface:
\begin{equation}
	E_y = \rho_s J,
	\label{eq:Ey}
\end{equation}
where $E_y$ and $J$ are the longitudinal components of the electric field strength and surface current density, $\rho_s$ is the surface resistance of the carbon nanotube~\cite{Slepyan1999}:
\begin{equation}
	\rho_s=j\frac{\pi^2 a \hbar (\omega-j\nu)}{2 e^2 \nu_F},
	\label{eq:rho_s}
\end{equation}
where $\nu_F$ is the Fermi velocity, $\omega$ is the cyclic frequency, $e$ is the electron charge, $\hbar$ is the Planck constant, $\nu$ is the relaxation frequency, $a$ is the nanotube radius, which is expressed through chirality indices $(m, n)$ as:
\begin{equation*}
	a=\frac{\sqrt{3}}{2\pi}d_0\sqrt{n^2+m^2+n m},
\end{equation*}
where $d_0=0.142$~nm is the distance between adjacent carbon atoms in the graphite plane.

Equation~\eqref{eq:rho_s} is applicable for metallic SWCNTs. It was obtained by G.\,Y.~Slepyan~\cite{Slepyan1999} in the approximation of the Boltzmann relaxation equation. The derivation of this equation with a generalization for the case of semiconductor nanotubes is given in the Appendix. It should be noted that an additional factor $\theta(g)$~\eqref{eq:current_theta} appears in the denominator~\eqref{eq:rho_s} for semiconductor SWCNTs. For instance, for semiconductor SWCNTs with radii $a \in [2.6\text{~\r{A} (6, 1)}$, $2.68\text{~nm (40, 39)}]$, $\theta(g)$ varies within the range $[3.16\cdot 10^{-9}$, $0.05]$.

Equation~\eqref{eq:rho_s} corresponds to the sum of the successive effective inductive and active resistances. The equation for active resistance contains a phenomenological parameter $\nu$ and requires further discussion. Due to the use of the Boltzmann relaxation equation in~Ref.~\cite{Slepyan1999}, the active resistance, similar to the Drude model~\cite{Drude1900}, is proportional to the inverse of the relaxation time. The parameter $\nu$ in~Ref.~\cite{Slepyan1999} is equal to $\nu_0=3.33\cdot10^{11}$~Hz.

Metallic SWCNTs are one-dimensional quantum conductors with ballistic conductivity.~\cite{Landauer1970, Frank1998, Kong2001} This is a phenomenon in which the scattering of charges occurs only at the ends of the conductor, whereas the conductor conductivity is constant and depends on neither the length nor the radius of the conductor. The SWCNT band structure is determined by two Dirac cones~\cite{Samsonidze2003, Ando2005}, therefore, in the ballistic mode, the SWCNT conductivity is equal to two quanta of conductivity~\cite{Frank1998, Kong2001} (the value of one quantum is equal to $e^2/(\pi\hbar)$). However, a resistance of the order of $\pi\hbar/(2e^2)\approx6.5\mathrm{~k\Omega}$ is observed for SWCNTs with a length significantly less than $1\mathrm{~\mu m}$ and increases essentially for long nanotubes. For instance, in Ref.~\cite{Li2004}, the SWCNT resistance was measured at direct current. In particular, a metallic SWCNT with a length $l=25\mathrm{~\mu m}$ has a resistance $R=180\mathrm{~k\Omega}$ at room temperature. If we use the experimental results~\cite{Li2004}, then for nanotubes with a length of $l=10-30\mathrm{~\mu m}$ for the relaxation frequency of metallic SWCNTs we obtain the following:
\begin{equation*}
	\nu = \frac{4e^2\nu_F}{\pi\hbar}\frac{R}{l}=1.81\cdot 10^{12}\ldots 1.95\cdot 10^{12}~\text{~Hz}.
\end{equation*}
Therefore, the value of $\nu_0$ in Ref.~\cite{Slepyan1999} seems for us to be underestimated.

A random sample of SWCNTs contains approximately a third of metallic ones. According to our estimate (see Appendix equation~\eqref{eq:current_mod3}), in the terahertz range the resistance of semiconductor nanotubes is several orders of magnitude larger. Therefore, the two-dimensional periodic grating contains only metallic SWCNTs in the layered heterostructures considered in this article.

\section{Theory}

\begin{figure}[!t]
	\centering
	\includegraphics{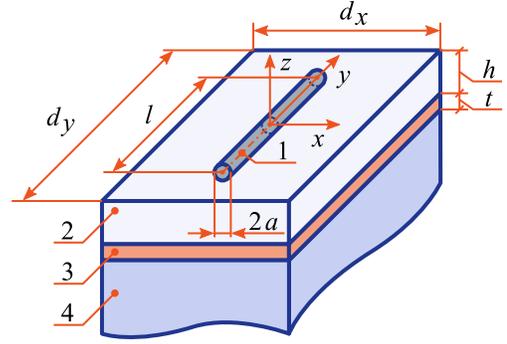}
	\caption{\label{fig:model} Geometric model of a frequency-selective surface: 1~--- one unit cell of the two-dimensional periodic SWCNT grating, 2~--- dielectric layer, 3~--- metallic layer, 4~--- semi-infinite dielectric substrate.}
\end{figure}

In this paper, a study of diffraction on heterostructures of two types with two-dimensional periodic grating is presented. The unit cell of the first structure, shown in Fig.~\ref{fig:model}, consists of SWCNT 1 with radius $a$ and length $l$ placed on a dielectric layer 2 with thickness $h$ and metallic layer 3 with thickness $t$ deposited on the surface of a semi-infinite dielectric substrate. The introduced Cartesian rectangular coordinate system has its origin located in the middle of the SWCNT; the z-axis is directed along the normal to the substrate surface and the y-axis is oriented along the nanotube symmetry axis. In the second structure, there is no metallic layer ($t=0$). 

Several SWCNTs can be placed parallel to each other in one unit cell in order to estimate the amplitude-frequency characteristics in the case of random distribution of SWCNT sizes. The geometrical parameters of each SWCNT and the distance between them are determined using a random number generator.

The electric field generated by currents $\vec J$ flowing on the nanotube surface $S$ is determined as follows:
\begin{equation}\begin{aligned}
	&\vec E(x,y,z) = -\frac{j}{\varepsilon_0 \varepsilon \omega}(\vec\nabla\vec\nabla+k^2)\cdot\\
	&\cdot\int_S\hat G(x, y, z;x',y',z')\vec J(x',y',z') \, dx' dy' dz',
\end{aligned}
\label{eq:E-field}
\end{equation}
where $\hat G$ is the tensor Green's function. For a multi-layered media, it is defined by Ref.~\cite{Lerer2021}. The SWCNT length is three orders of magnitude larger than its radius; therefore, only the longitudinal component of the surface current $\vec J = J\vec e_y$ can be taken into account in SWCNTs.

Substituting \eqref{eq:E-field} into \eqref{eq:Ey}, the integro-differential equation is obtained:
\begin{equation}\begin{aligned}
	E_y^{\mathrm{ext}}(x,y,z) &- \frac{j}{\varepsilon_0 \varepsilon \omega}\int_S\left(k^2-\frac{\partial^2}{\partial y \partial y'}\right)\cdot\\
	&\cdot G_{yy}(x, y, z;x',y',z')\cdot\\
	&\cdot J(x',y',z') \, dx' dy' dz'=\rho_s J(x,y,z),
\end{aligned}
\label{eq:ide}
\end{equation}
where the observation points in the cylindrical coordinate system ($x=a \cos \varphi$, $y$, $z=a \sin \varphi$) and the source points ($x'=a \cos \varphi'$, $y'$, $z'=a \sin \varphi'$) are located on the SWCNT surface ($0\le\varphi<2\pi$, $|y|\le l$); $\vec E^{\mathrm{ext}}$ is the external electromagnetic field (the field in a multilayer structure without SWCNT taken into account when a plane electromagnetic wave is incident). Obtaining $\vec E^{\mathrm{ext}}$ is a simple issue.~\cite{Lerer2012}

For a two-dimensional periodic grating of SWCNTs placed on the surface of a multilayered substrate, with periods $d_x$ and $d_y$, the Green's function can be written as a series. The component $\hat G_{yy}$ of the tensor Green's function has the following form:
\begin{equation*}\hspace*{-0.25em}
	G_{yy} = \frac{1}{d_x d_y}\sum_{\substack{m=-\infty,\\n=-\infty}}^\infty\hspace*{-0.75em}\tilde G_{yy;mn}(z,z')\exp[j\alpha_m (x - x') + j\beta_n (y-y')],
\end{equation*}
where $\alpha_m=2\pi m / d_x + k_x$, $\beta_n=2\pi n / d_y + k_y$, $k_x$ and $k_y$ are the projections of the wave vector of the incident wave, the function $\tilde G_{yy;mn}$ in each of the layer satisfies the equation:
\begin{equation}
	\hspace*{-1em}\left[\frac{\partial^2}{\partial z^2}-\alpha_m^2-\beta_n^2+k^2\varepsilon(z)\right] \tilde G_{yy;mn}(z,z') = - \delta(z-z'),
	\label{eq:condition}
\end{equation}
where $\varepsilon(z)$ is constant within each layer, $\delta$ is Dirac' delta function. It should be noted that the transverse component of the current can be neglected, since the SWCNT length is several times larger than its diameter.

In Ref.~\cite{Lerer2012}, the tensor Green's function $\tilde{G}_{nm}$ was constructed satisfying Equation~\eqref{eq:condition} and the continuity condition for tangential components of the electric field strength. The recurrence equations for $\tilde G_{mn}$ in Ref.~\cite{Lerer2012} can be used for an arbitrary number of layers and are protected from overflow.

In our study the SWCNTs are located on the surface of a multilayered substrate. In this case, the solution~\eqref{eq:condition} in the upper layer is the function:
\begin{equation*}
	\tilde G_{yy;mn}(z,z') = \tilde G_{0;mn}\exp\left(-\sqrt{\alpha_m^2+\beta_n^2-k^2}|z-z'|\right).
\end{equation*}
The algorithm proposed in Ref.~\cite{Lerer2012} is used to determine the constants $\tilde G_{0;mn}$.

Equation~\eqref{eq:ide} is solved by the Bubnov-Galerkin projection method. The current density function $\vec J$ is approximated by a linear combination of the basis functions:
\begin{equation*}
	J_y(\varphi, y) = \frac{1}{2\pi a}\sum_{v'=0}^N \frac{Y_{v'}}{\hat Y_{v'}} C_{v'}^{(3/2)}\left(\frac{y}{l}\right)\left[1-\frac{y^2}{l^2}\right],
\end{equation*}
where $C_\nu^{(\alpha)}$ is Gegenbauer polynomial (ultraspherical polynomial), $Y_{v'}$ is unknown coefficients, the values $\hat Y_{v'}$ are selected so that the condition is satisfied:
\begin{equation*}
	\frac{1}{\hat Y_{v'}} \int\limits_{-l}^l C_{v'}^{(3/2)}\left(\frac{y}{l}\right)\left[1-\frac{y^2}{l^2}\right] \exp (j k_y y) \, dy = \frac{J_{v'+3/2}(k_y l)}{(k_y l)^{3/2}},
\end{equation*}
where $J_\nu$ is Bessel function of the first kind.

Using the projection method, a system of linear equations of $N+1$ order with respect to the unknowns $Y_{v'}$ is obtained. The matrix elements in the system of linear equations contain a double integral over the SWCNT surface, which is calculated numerically. The left side of equation~\eqref{eq:E-field} is used to find the reflected and transmitted waves.

A system of linear equations has good convergence. Usually, it is sufficient to solve a fifth-order system for calculations with an internal convergence error of 0.5\%. The time of numerical calculation at a given frequency is less than 0.5s on 1 core of Intel Core i3-2100CPU@3.10GHz.

\section{Results and Discussion}\label{sec:results}

Below we present the results of a computational study of the diffraction of electromagnetic waves with $s$-polarization (the electric field is directed along the SWCNT) and $p$-polarization (the electric field is directed across the SWCNT) for two types of frequency-selective surfaces shown in Figure~\ref{fig:model}.

Both types of FSS contains the SWCNTs with chirality indices (31, 31) and the radius $a=2.102$~nm, which form a two-dimensional periodic grating with periods $d_x=0.1\mathrm{~\mu m}$ and $d_y=25\mathrm{~\mu m}$. The nanotubes have a length $l=20\mathrm{~\mu m}$. The SWCNT relaxation frequency is $\nu=\nu_0=3.33\cdot 10^{11}$~Hz. Zinc Oxide with a refractive index $n=1.96$ is selected as the material of the semi-infinite dielectric substrate 4.

The first structure (FSS1) is ideally transmissive with a minimum transmission coefficient $T$ at wave resonance with $s$-polarization and $T=100\%$ for a wave with $p$-polarization. There is no metallic layer 3 in FSS1. In order to realize an ideal quarter-wavelength coating, the dielectric layer 2 with a refractive index $n=1.4\approx\sqrt{1.96}$ (approximately equal to that in SiO\textsubscript{2}) is used. The layer has a thickness $h=109.3\mathrm{~\mu m}$ that approximately equals to a quarter of the wavelength for the insulation material at the resonant frequency.

The second structure (FSS2) is ideally reflective with a minimum reflection coefficient $R$ at wave resonance with $s$-polarization and $R=100\%$ for a wave with $p$-polarization. Metallic layer Au with a thickness $t=0.1\mathrm{~\mu m}$ and conductivity similar to that of a bulk material was selected as a reflective metallic film 3. The choice of dielectric layer 2 is similar to the case of FSS1 except for a slightly different refractive index $n=1.45$.

It should be noted that for FSS2 structure, instead of a metallic mirror, it is possible to realize a dielectric mirror, consisting of several alternating quarter-wave dielectric layers.

\begin{figure}[!t]
	\centering
	\includegraphics{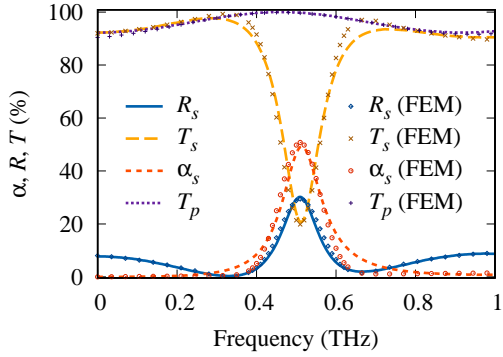}
	\caption{\label{fig:afc_fss1} Amplitude-frequency characteristics of the power  reflection $R$, transmission $T$, and absorption $\alpha$ coefficients for FSS1 heterostructure when $s$- and $p$-polarization are considered.}
\end{figure}

Figure~\ref{fig:afc_fss1} shows the amplitude-frequency characteristics of the power reflection $R$ and transmission $T$ coefficients for FSS1, whereas $\alpha$ is the absorption coefficient. The $s$ and $p$ indices denote  the type of wave polarization. The circular resonant frequency is $1.5$ times greater than $\nu_0$.

In the resonance region, a $p$-polarized wave passes through the grating without reflection. The transmission coefficient of a $s$-polarized wave may be less than $20\%$. It should be noted that the length of an equivalent metallic half-wave oscillator at a frequency $0.5$~THz is approximately $300\mathrm{~\mu m}$, which is $15$ times greater than for SWCNTs.

\begin{figure}[!t]
	\centering
	\includegraphics{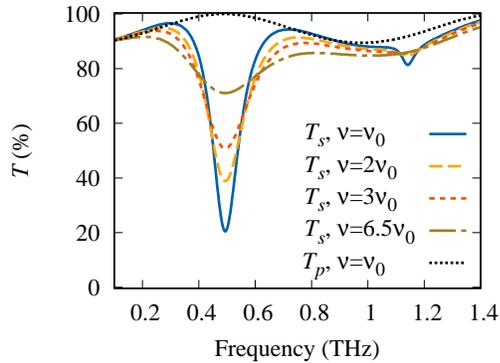}
	\caption{\label{fig:fdtc_fss1} Frequency dependence of the transmission coefficient $T$ for FSS1 heterostructure at different SWCNT relaxation frequencies.}
\end{figure}

\begin{figure}[!t]
	\centering
	\includegraphics{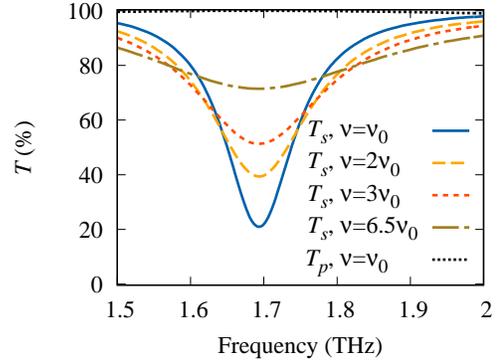}
	\caption{\label{fig:fdtc_fss1_shrot} Frequency dependence of the transmission coefficient $T$ for FSS1 heterostructure consisting of short SWCNTs at different SWCNT relaxation frequencies.}
\end{figure}

The paper further considers the effect of losses on the frequency characteristics of the reflection coefficient for FSS1 heterostructure (Fig.~\ref{fig:fdtc_fss1}, \ref{fig:fdtc_fss1_shrot}). Similar to a simple oscillator, increasing the loss ($\nu$) extends the resonant curve without changing the resonant frequency. Thus, if our estimate $\nu$ is correct, then the resonance on long SWCNTs is fairly well pronounced at room temperature.

Figure~\ref{fig:fdtc_fss1_shrot} shows the frequency characteristics of the transmission coefficient $T$ for FSS1 with SWCNT length $l=3\mathrm{~\mu m}$ and unit cell periods $d_x=0.1\mathrm{~\mu m}$, $d_x=2.5\mathrm{~\mu m}$. Since SWCNT surface resistance depends on the frequency, the scalability principle is not observed. Shortening SWCNTs by ten times (from $20$ to $2\mathrm{~\mu m}$) increases the resonance frequency by approximately $3.5$ times. The circular resonant frequency is $5$ times greater than $\nu_0$. The width of the resonance curve almost remains unchanged. Due to the higher frequency, the Q-factor of the resonance is higher for a short SWCNTs.

\begin{figure}[!t]
	\centering
	\includegraphics{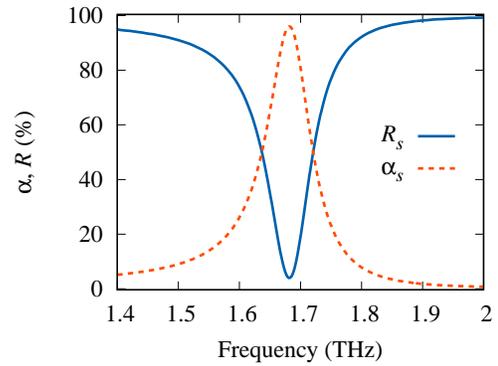}
	\caption{\label{fig:fd_ra_fss2_dl} Frequency dependence of the reflection $R$ and absorption $\alpha$ coefficients for FSS2 heterostructure at the dielectric layer thickness $h=153.4\mathrm{~\mu m}$.}
\end{figure}

\begin{figure}[!t]
	\centering
	\includegraphics{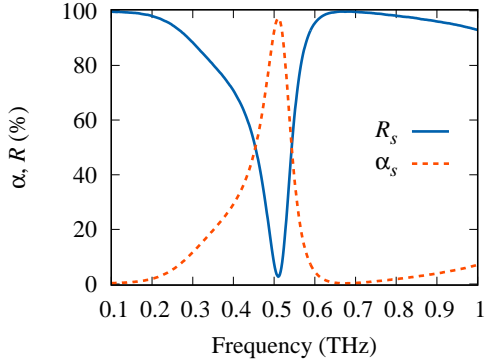}
	\caption{\label{fig:fd_ra_fss2_ll} Frequency dependence of the reflection $R$ and absorption $\alpha$ coefficients for FSS2 heterostructure for the lattice period $d_x=0.1\mathrm{~\mu m}$ $d_y=2.5\mathrm{~\mu m}$ and SWCNT length $l=2\mathrm{~\mu m}$. The thickness of the dielectric layer equals $h=47\mathrm{~\mu m}$.}
\end{figure}

\begin{figure}[!t]
	\centering
	\includegraphics{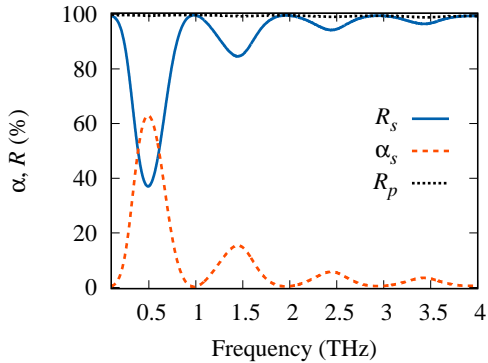}
	\caption{\label{fig:fd_ra_fss2_nu} Frequency dependence of the reflection $R$ and absorption $\alpha$ coefficients for FSS2 heterostructure at the relaxation frequency $\nu = 10 \nu_0$.}
\end{figure}

Figures~\ref{fig:fd_ra_fss2_dl}--\ref{fig:fd_ra_fss2_nu} show the amplitude-frequency characteristics of FSS2 structure. When the incident electromagnetic wave is $p$-polarized, the grating is an almost perfect mirror. In the resonance region of the $s$-polarized wave, the losses in both the SWCNT and the metallic film increase significantly. When the thickness of the dielectric layer is equal to $0.25$--$0.375$ of the wavelength, the incident power on the SWCNT and the reflected one from the metallic layer (Au) are added in phase at the resonant frequency. The higher resonances are poorly pronounced.

With an increase in the relaxation frequency $\nu$, the level of the main resonance decreases and higher resonances appear (Fig.~\ref{fig:fd_ra_fss2_nu}). They are due to the large reflection coefficient (from SWCNTs) of standing electromagnetic wave in the dielectric layer.

To estimate the amplitude-frequency characteristics in the case of a random distribution of SWCNT sizes, it is assumed that the unit cell contains an $M$ number of SWCNTs. The geometrical parameters of each SWCNT and the distance between them are determined by a random number generator. The following parameters are additionally set for calculations of such structure:
\begin{itemize}
	\item average distance between nanotubes $d$ ($\mathrm{\mu m}$) and its maximum deviation $\Delta d$ ($\mathrm{\mu m}$),
	\item maximum deviation of the nanotube radii $\Delta a$ ($\mathrm{\mu m}$),
	\item maximum deviation of the nanotube length $\Delta l$ ($\mathrm{\mu m}$),
	\item maximum deviation of the angles of inclination of nanotubes to the $x$-axis.
\end{itemize}

\begin{figure}[!t]
	\centering
	\includegraphics{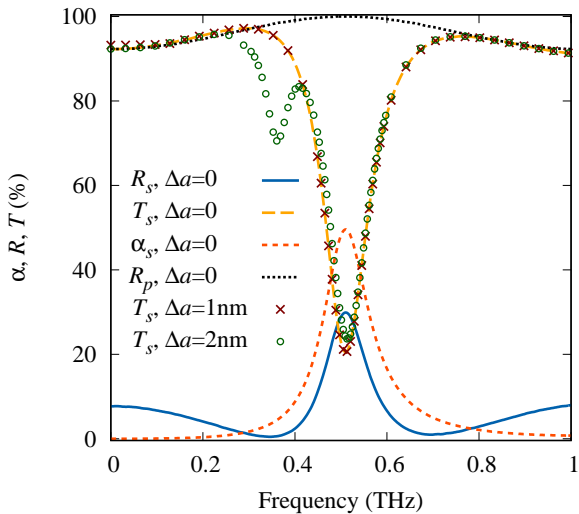}
	\caption{\label{fig:afd_rnd_fss1} Amplitude-frequency dependence of the FSS1 heterostructure with a random scatter of SWCNT sizes.}
\end{figure}

Figure~\ref{fig:afd_rnd_fss1} shows the amplitude-frequency dependence of the FSS1 heterostructure containing identical SWCNTs (curves with $\Delta a = 0$) and
those characterizing by a random distribution of geometrical parameters. A small deviation from the average value of the SWCNT radii ($l=20\mathrm{~\mu m}$, $\Delta a=1$~nm, red cross-shaped markers) has an extremely weak effect on the resonance position. In the case of a scatter of SWCNT lengths ($l=20\mathrm{~\mu m}$, $\Delta a=2$~nm, green circle-shaped markers), the main resonance also changes weakly, and well-pronounced additional resonances appear. Their position and depth are highly dependent on the magnitude $\Delta l$. 

\begin{figure}[!t]
	\centering
	\includegraphics{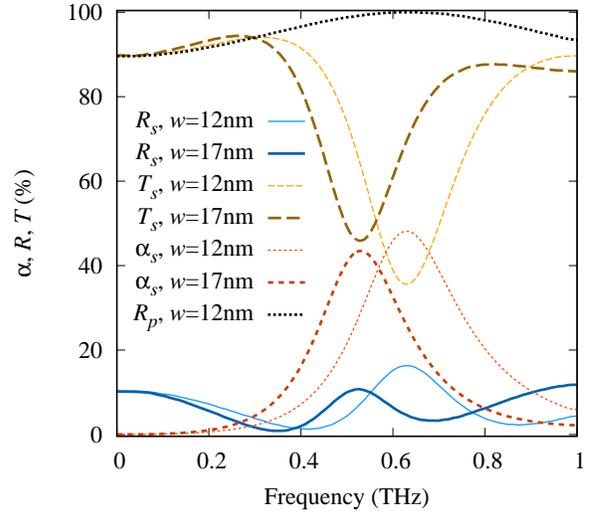}
	\caption{\label{fig:afd_graphene} Amplitude-frequency characteristics of a two-dimensional periodic grating of graphene strips with a width of $w=12$~nm and $w=17$~nm.}
\end{figure}

The paper further compares the characteristics of SWCNT FSS with the characteristics of similar FSS formed by graphene strips. The surface resistance of graphene can be described by the Kubo formula~\cite{Hanson2008}. This formula is also used when calculating $\mathrm{\mu m}$-sized graphene strips. At nanometer sizes, size effects shall occur.

When graphene is rolled up into a nanotube, size effects also occur, and the Kubo formula for SWCNTs may not be applicable. However, it should be expected that the resonant frequencies lie in the same frequency range. To verify this statement, FSS1 and a similar heterostructure were calculated, where graphene strips~\cite{Lerer2018} are used instead of nanotubes. The lengths of strips and SWCNTs are the same and equal to $20\mathrm{~\mu m}$. The width of strip $w$ is equal to the diameter of the nanotube $2a$ ($\mu = 0.25$~eV, $\tau = 1$~ps, $T=300^\circ$~K). The resonances of FSS based on graphene strips depend on the relaxation time $\tau$ and chemical potential $\mu$, which can be controlled by applying gate voltage and chemical doping. At room temperature $T=300^\circ$~K ($\mu = 0.25$~eV, $\tau = 1$~ps), these resonances lie in the same frequency range as FSS resonances from SWCNTs.

Figure~\ref{fig:afd_graphene} shows the amplitude-frequency characteristics of a two-dimensional periodic grating of graphene strips with a width of $w=12$~nm and $w=17$~nm. Since SWCNTs are located in the air and the graphene strip lies on an insulation material, it can be assumed that the above estimate of the strip width is overestimated. The resonant frequency of this grating almost completely coincides with the resonance frequency of the CNT grating (Fig.~\ref{fig:afc_fss1}), whereas the frequency characteristics are close to the FSS1 characteristics from CNTs at $\nu=2\nu_0$.

\section {Conclusion}\label{sec:conclusion}

The theory of constructing the tensor Green's function for Maxwell's equations solved by the Bubnov-Galerkin projection method was developed for the numerical electrodynamic simulation of electromagnetic wave diffraction on a two-dimensional periodic grating of SWCNTs and graphene strips located on a multilayered substrate. On the basis of the Bubnov-Galerkin method, algorithms for numerical calculation, and software were developed for mathematical modeling of electromagnetic wave diffraction on heterostructures with two-dimensional periodic SWCNT grating. The presence of a surface plasmon-polariton resonance on SWCNTs in the terahertz range was shown. The amplitude-frequency dependence of resonant-reflecting and resonant-transmitting frequency-selective surfaces was calculated. The model of diffraction grating with a random scatter of SWCNT sizes was studied. The frequency characteristics of frequency-selective surfaces for SWCNTs and graphene strips were compared.

\appendices

\section{Extended Slepyan-Hanson theory}

\begin{figure*}[!t]
	\centering
	\includegraphics{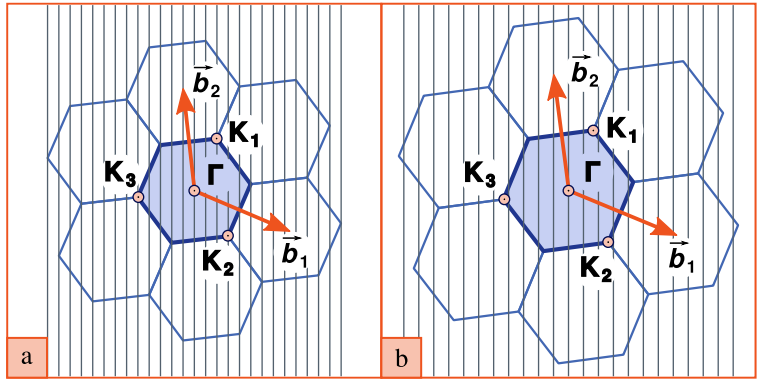}
	\caption{\label{fig:tube_chirallity} The first and adjacent Brillouin zones in reciprocal space of graphene with superimposed cutting lines: (a)~~--- metallic SWCNT, (b)~--- semiconductor SWCNT. The vertices of hexagons correspond to the points $\mathbf{K}$, where the electron dispersion is characterized by a Dirac cone and a constant group velocity $\nu_F$. Each point $\mathbf{K}$ is divided between three adjacent zones, so the first Brillouin zone has two non-equivalent points $\mathbf{K}$. The marked points $\mathbf{K_1}$, $\mathbf{K_2}$, and $\mathbf{K_3}$ are equivalent with respect to the translations of the reciprocal space.}
\end{figure*}

The reciprocal space of any SWCNT is a set of parallel equidistant cutting lines lying within the first Brillouin zone (FBZ) of graphene~\cite{Samsonidze2003} (Fig.~\ref{fig:tube_chirallity}). The cutting lines are numbered by index $s$ and directed along the axis of the nanotube. Their number is equal to the number of hexagons per SWCNT period, and the distance between lines equals $2\pi/P=1/a$, where $P$ is the nanotube perimeter. The equilibrium Fermi-Dirac distribution for electrons in SWCNTs is as follows~\cite{Ando2005}:
\begin{equation*}
	\eta(E_s, T) = \frac{\displaystyle 2}{\displaystyle 1+\exp\left(\frac{E_s(k)-\mu}{k_B T}\right)},
\end{equation*}
where $E_s(k)$ is the dependence of the electron energy on the wave vector for the cutting line with index $s$, $k_B=1.38\cdot 10^{-23}\mathrm{~J/K}$ is the Boltzmann constant, $T$ is the absolute temperature, and the chemical potential $\mu$ in the issue under consideration is zero. The factor $2$ appears in the numerator, since two electrons per graphene unit cell participate in the formation of the band structure, and, accordingly, the state $E_s(k)$ simultaneously corresponds to two different spin orientations.

The paper further considers the non-equilibrium electron distribution $\eta_c$. This distribution occurs in an external alternating electric field. Unlike the usual Fermi—Dirac distribution $\eta$, it depends on time $t$ and coordinate $y$ directed along the nanotube. Following Ref.~\cite{Slepyan1999}, the Boltzmann relaxation equation can be written as follows:
\begin{equation}
\begin{aligned}
	\frac{d}{dt}\eta_c&=\frac{\partial \eta_c}{\partial t}+\frac{\partial \eta_c}{\partial p_s}\frac{\partial p_s}{\partial t}+\frac{\partial \eta_c}{\partial y}\frac{\partial y}{\partial t}=\\
	&=\frac{\partial \eta_c}{\partial t}+e E_y \frac{\partial \eta_c}{\partial p_s} + V_y \frac{\partial \eta_c}{\partial y}=\nu(\eta-\eta_c),
\end{aligned}	
\label{eq:boltzmann}
\end{equation}

where the relaxation processes are taken into account in the right part of the equation; the electric field $E_y$ is directed along the SWCNT axis, $e E_y$ is the force that changes the quasimomentum of the electron $p_s$, $V_y$ is its velocity, $k$ is the wave number of the electron, $\nu$ is the relaxation frequency.

Rewriting $\eta_c$ in \eqref{eq:boltzmann} as the sum of the equilibrium $\eta$ and non-equilibrium distribution $\delta\eta$
\begin{equation*}
	\eta_c = \delta\eta(t, y)+\eta,
\end{equation*}
we obtain the following differential equation:
\begin{equation}
	\frac{\partial \delta\eta}{\partial t} + e E_y(t)\frac{\partial \eta}{\partial p_s} + V_y \frac{\partial \delta\eta}{\partial y} = -\nu\delta\eta.
	\label{eq:de_eta}
\end{equation}

It is assumed that the electric field changes in time according to the harmonic law along $y$-axis. Substituting into~\eqref{eq:de_eta} the expressions $E_y(t)=E_0\exp\left(j(\omega t - k y)\right)$, $\delta\eta(t)=\delta\eta_0\exp\left(j(\omega t - k y)\right)$, and $V_y(t)=V_{y\,0}\exp\left(j(\omega t - k y)\right)$, the following equation is obtained
\begin{equation}
	\delta\eta_0\left(\nu+j \omega+j k V_{y\,0}\right) = - e E_0 \frac{\partial \eta}{\partial p_s}.
	\label{eq:de_eta_mod}
\end{equation}
Rearrangement of the equation~\eqref{eq:de_eta_mod} leads to the following:
\begin{equation}
	\delta\eta_0=\frac{j e E_0}{\omega - k V_{y\,0} - j \nu}\frac{\partial \eta}{\partial p_s}.
\end{equation}

In the issue under consideration, the value $V_{y\,0}$ is limited by the Fermi velocity $\nu_F$ of electrons in graphene. The contribution of the small term $k V_{y\,0}$ in the denominator can be neglected~\cite{Hanson2008}. The amplitude of the total electron current $I_{e0}$ can only be related to the amplitude of the non-equilibrium part of the distribution $\delta\eta_0$:
\begin{equation}
	I_{e0}=\frac{1}{l}\sum_{s, k} e \delta\eta_0 \frac{\partial E_s(k)}{\hbar\partial k},
	\label{eq:current}
\end{equation}
where $\hbar^{-1}\partial E_s(k)/\partial k = V_g(k)$ is the electron group velocity and $l$ is the SWCNT length. Taking into account in~\eqref{eq:current} that
\begin{equation*}
	\frac{\partial \eta}{\partial p_s} = \frac{\partial \eta}{\partial E_s} \frac{\partial E_s}{\partial p_s}
	=\frac{\partial \eta}{\partial E_s} \frac{\partial E_s(k)}{\hbar \partial k}
\end{equation*}
we obtain the following:
\begin{equation*}
	I_{e0}=\frac{1}{l}\frac{j e^2 E_0}{\omega-j \nu}\sum_{s, k}\frac{\partial \eta}{\partial E_s}\left[\frac{\partial E_s(k)}{\hbar\partial k}\right]^2.
\end{equation*}

The valence band in SWCNTs is symmetric with the conduction band. Assuming the relaxation time for holes is the same as for electrons, and taking into account the hole current in the valence band (equal to $I_{e0}$), we use the integration over the wave number and rewrite the amplitude of the total electron and hole current as:
\begin{equation}
	I_0=\frac{j}{\pi}\frac{e^2 E_0^2}{\omega-j \nu}\sum_{s}\int\limits_{k_{s\,1}}^{k_{s\,2}}\frac{\partial \eta}{\partial E_s}\left[\frac{\partial E_s(k)}{\hbar\partial k}\right]^2\,d k,
	\label{eq:current_mod1}
\end{equation}
where the points $k_{s\,1}$ and $k_{s\,2}$ lie at the intersection of the cutting line with $s$ index with FBZ boundaries.

The main contribution to~\eqref{eq:current_mod1} for metallic nanotubes is caused by a pair of cutting lines passing through two translationally nonequivalent points $\mathbf{K}$ of the first Brillouin zone of graphene, commonly referred to as the conduction channels. For semiconductor tubes, this is a pair of cutting lines that are as close as possible to $\mathbf{K}$ points. Since the main contribution to~\eqref{eq:current_mod1} is made only in the vicinity of the points $\mathbf{K}$, the integration can be extended to infinity and the origin can be chosen at one of the points $\mathbf{K}$:
\begin{equation}
	I_0=-\frac{4}{\pi}\frac{j e^2 E_0^2}{k_B T (\omega - j \nu)}\int\limits_{-\infty}^\infty \frac{\left[\frac{\partial E_s(k)}{\hbar \partial k}\right]^2 \exp \left(\frac{E_s(k)}{k_B T}\right)}{\left[1+\exp\left(\frac{E_s(k)}{k_B T}\right)\right]^2}\,d k
	\label{eq:current_mod2}
\end{equation}
Factor 4 in~\eqref{eq:current_mod2} is obtained by taking into account two nonequivalent points $\mathbf{K}$ and factor 2 in the Fermi–Dirac distribution function.

For \textit{semiconductor nanotubes}, the nearest cutting line passes in the reciprocal space at a distance $\Delta k = (3a)^{-1}$ from the point $\mathbf{K}$. For the cone section describing the shape of the conduction band around the point $\mathbf{K}$, $E_s(k)=\nu_F\hbar\sqrt{k^2-(\Delta k)^2}$. The substitution of $\chi=\hbar k \nu_F / (k_B T)$ in~\eqref{eq:current_mod2} leads to the following
\begin{equation}
	I_0=-\frac{4j}{\pi\hbar}\frac{e^2 E_0 \nu_F}{\omega-j \nu}\theta(g),
	\label{eq:current_mod3}
\end{equation}
where
\begin{align}
	\theta(g)=&2\int\limits_0^\infty \frac{x^2 \exp\left(\sqrt{x^2+g^2}\right)}{\left(x^2+g^2\right)\left(1+\exp\left(\sqrt{x^2+g^2}\right)\right)^2}\,dx,
	\label{eq:current_theta}\\
	g=&\frac{\hbar\nu_F}{3 a k_B T}.\nonumber
\end{align}

Taking into account that $\nu_F=\sqrt{3}\gamma_0 b_g / (2\hbar)$, $b_g\approx\sqrt{3}d_0$ is the graphene lattice parameter, and $\gamma_0\approx3$eV is the strong coupling constant in graphene, the parameter $g$ can be expressed as $g=\gamma_0 b_g / (2\sqrt{3}a k_B T)$. However, it is more convenient to write the parameter $g$ in terms of the chiral indices $(n, m)$ of a semiconductor nanotube, where $n-m$ is not a multiple of 3:
\begin{equation*}
	g = \frac{\pi \gamma_0}{\sqrt{3} k_B T \sqrt{n^2+m^2+nm}}.
\end{equation*}

For thick semiconductor nanotubes, a cutting line at a distance $\Delta k = 2a/3$ can be also considered. It is advisable if the condition $\theta(2g)\ll\theta(g)$ is violated. Then in~\eqref{eq:current_mod3}, the factor $\theta(g)$ should be substituted by $\theta(g)+\theta(2g)$. It should be noted that the factor $\theta(g)$ significantly increases the resistance of semiconductor nanotubes. For instance, at room temperature for a thin semiconductor nanotube $(6,1)$, we obtain $\theta(g)=\theta(18.97)\approx3.16\cdot 10^{-9}$.

For \textit{metallic nanotubes} $\Delta k = 0$ and $\theta(0) = 1$, the equation~\eqref{eq:current_mod3} is reduced the well-known form~\cite{Hanson2008}:
\begin{equation}
	I_0=-\frac{4j}{\pi\hbar}\frac{e^2 E_0 \nu_F}{\omega-j \nu}.
	\label{eq:current_mod4}
\end{equation}

Equation~\eqref{eq:rho_s} can be obtained from formula~\eqref{eq:current_mod4} if the following is taken into consideration:
\begin{equation*}
	\rho_s=\frac{E_0}{J},\quad J = \frac{I_0}{P}.
\end{equation*}

In conclusion of this section, the applicability range of the considered theory is discussed. The high frequency limitation appears due to fundamental absorption because of electron transitions to the conduction band. In the vicinity of the cone circumscribing the shape of the band structure around the point $\mathbf{K}$, the distance (in energy) between the conduction band and the valence band is defined as $\omega=2\hbar\nu_F k$, where the wave-vector length $k$ is calculated as a distance from the point $\mathbf{K}$. The nearest cutting line that does not pass through the point $\mathbf{K}$ in the reciprocal space of the metal nanotube corresponds to $k=\hbar^{-1}a^{-1}$. Then the absorption does not occur if $\omega<2\nu_F a^{-1}$.

As an example, consider nanotubes with a radius $a=2.102$~nm and chirality indices (31, 31). For this case the limiting frequency equals $f_c=1.47\cdot 10^{14}$~Hz ($\lambda_c=2039$~nm). If the heterostructure contains semiconductor tubes with close radii, they begin to absorb resonantly at a frequency approximately 3 times lower. Nanotubes with smaller radii absorb at higher frequencies. Therefore, the applicability region of the theory is limited to the frequency $f_c\approx 10^{14}$~Hz.

\section*{Data Availability}

The data that support the findings of this study are available from the corresponding author upon reasonable request.

\section*{Acknowledgments}
	Lerer~A.\,L. and Rochal~S.\,B. acknowledge financial support from the Russian Foundation for Basic Research (Grant No. 18-29-19043~mk).

\ifCLASSOPTIONcaptionsoff
  \newpage
\fi

\bibliographystyle{IEEEtran}
\bibliography{timoshenko}

\newpage

\begin{IEEEbiography}[{\includegraphics[width=1in,height=1.25in,clip,keepaspectratio]{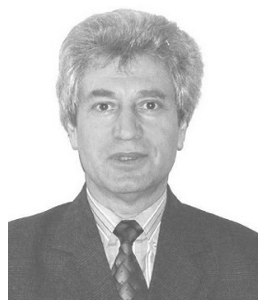}}]{Lerer Alexander Mikhailovich} is a Professor of Applied Electrodynamics and Numerical Modeling at the Faculty of Physics of Southern Federal University, Russia.
	
His research interests are in computational electrodynamics, modeling of diffraction and propagation of monochromatic and non-stationary electromagnetic waves, nanostructures in the millimeter and optical bands.
\end{IEEEbiography}

\begin{IEEEbiography}[{\includegraphics[width=1in,height=1.25in,clip,keepaspectratio]{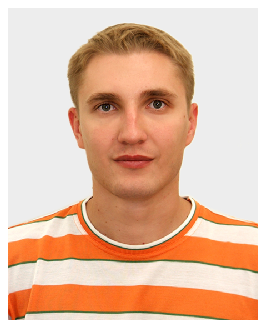}}]{Timoshenko, Pavel Evgenjevich} received his PhD degree in Physical and Mathematical Sciences from Southern Federal University, Russia. 
	
Since 2006, he has been with the Faculty of Physics of Southern Federal University, where he became an assistant professor in 2018. His current research interests are in the fields of computational electrodynamics and acoustics, numerical modeling of acoustics and optics phenomena and devices, periodic structures (phononic and photonic crystals).
\end{IEEEbiography}

\begin{IEEEbiography}[{\includegraphics[width=1in,height=1.25in,clip,keepaspectratio]{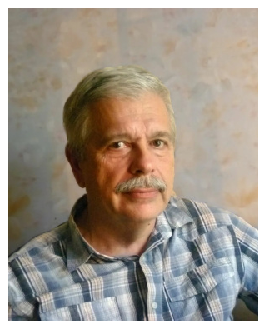}}]{Rochal Sergei Bernardovich} received the Doctor of Physical and Mathematical Sciences degree in physics of condensed matter from Rostov State University, Rostov-on-Don, Russia.
	
He is currently with the Faculty of Physics, Southern Federal University, Rostov-on-Don, Russia. His is the author or coauthor of more than 80 journal articles, referee in several European and American physical journal. His current research interests include physical properties of carbon nanotubes, self-assembly of biological nano-objects, theory of phase transitions, colloidal crystals, self-assembly and structural changes in the viral capsids.
\end{IEEEbiography}

\vfill

\end{document}